**Title:** Identification of $^1$H-NMR Spectra of Xyloglucan Oligosaccharides: A Comparative Study of Artificial Neural Networks and Bayesian Classification Using Nonparametric Density Estimation.

**Authors:** Faramarz Valafar, Homayoun Valafar, William S. York

**Conference:** IC-AI '99

**Paper identification number:** 260A

# Identification of $^1$H-NMR Spectra of Xyloglucan Oligosaccharides: A Comparative Study of Artificial Neural Networks and Bayesian Classification Using Nonparametric Density Estimation

Faramarz Valafar
University of Georgia, CCRC
Athens, GA 30602
Email:
faramarz@ccrc.uga.edu

Homayoun Valafar
University of Georgia, CCRC
Athens, GA 30602
Email:
homayoun@ccrc.uga.edu

William S. York
University of Georgia, CCRC
Athens, GA 30602
Email: will@ccrc.uga.edu

*Abstract:*

*Proton nuclear magnetic resonance ($^1$H-NMR) is a widely used tool for chemical structural analysis. However, $^1$H-NMR spectra suffer from natural aberrations that render computer assisted automated identification of these spectra difficult, and at times impossible. Previous efforts have successfully implemented instrument dependent, or conditional identification of these spectra. In this paper we report the first instrument independent computer assisted automated identification system for a group of complex carbohydrates known as the xyloglucan oligosaccharides. The developed system is also implemented on the world wide web (http://www.ccrc.uga.edu) as part of an identification package called the CCRC-Net, and is intended to recognize any submitted $^1$H-NMR spectrum of these structures with reasonable signal-to-noise ratio, recorded on any 500 MHz NMR instrument. The system uses Artificial Neural Networks (ANNs) technology and is insensitive to instrument and environment dependent variations in $^1$H-NMR spectroscopy. In this paper, comparative results of the ANN engine versus a multidimensional Bayes' classifier is also presented.*

## 1. Introduction:

**$^1$H-NMR spectroscopy.** NMR (nuclear magnetic resonance) spectroscopy is a widely used tool for chemical analysis. It is used to identify materials, determine the chemical structure of organic compounds, and can be used to quantify chemical substituents or the components of chemical mixtures. The proton (**$^1$H**) is the nuclide that is most frequently observed by NMR. When a sample is placed in a strong magnetic field, the magnetically active nuclei become aligned. The resulting sample magnetization can be manipulated by applying a very brief magnetic field pulse that oscillates at radio frequency (RF). Such RF pulses perturb the sample magnetization, which can be observed via its induction of a current in a detector coil as the magnetization relaxes back to its equilibrium state. The resulting "free induction decay" (FID) contains information regarding the chemical environment of nuclei within the chemical sample, and thus can be used to identify and quantitate individual chemical components of the sample. The FID consists of a mixture of sinusoidal oscillations in the time-domain with decaying amplitudes. The time-domain signal is normally transformed (usually using Fourier transform) into the frequency domain. Figure 1 (located at the end of this article), illustrates two examples of a frequency domain signal (spectrum) of a xyloglucan oligosaccharide.

$^1$H-NMR spectra, in general, suffer from environmental, instrumental, and other types of variations that manifest themselves in a variety of aberrations. Low signal-to-noise ratio [1, 2, 4], baseline drifts [3, 4, 7], frequency shifts due

to temperature variations, line broadening and negative peaks due to phasing problems, and malformed peaks (or overlapped peaks) due to inaccurate shimming, are among the most prominent and common aberrations. For example, Figure 1, shows two $^1$H-NMR spectra of a complex carbohydrate. The spectrum labeled (B) in this figure suffer from a variety of the above mentioned aberrations, and contamination by lactate, frequently introduced by touching laboratory glassware with bare hands. It is important to realize that this spectrum by no means represents a worst case scenario, and it does not represent the level of complexity present in the problem of instrument independent identification of $^1$H-NMR spectra of xyloglucan oligosaccharides. Spectrum (B) is merely a demonstration of some types of possible aberrations.

For the purpose of automated identification of these spectra, elimination of the above mentioned aberrations becomes essential, as they can lead to erroneous identification [1-7]. A variety of signal processing techniques have been applied to "clean up" $^1$H-NMR spectra. For instance, signal averaging[1] [4] and apodization[2] [4] have become standard ways of improving the signal-to-noise ratio. To correct baseline problems, a number of techniques have been used such as parametric modeling using a priori knowledge [3, 5], optimal associative memory (OAM) [5], and spectral derivatives [6]. Other mathematical techniques have also been introduced to address each specific type of aberration encountered in $^1$H-NMR spectra.

---

[1] In signal averaging a spectrum is recorded several times. Each recorded signal is referred to as a "transient." The final spectrum is the arithmetic average of all the transients. The hope is that by doing so the zero mean components of the noise present in the signal will be averaged out.

[2] Apodization is a type of low (high) pass filtering performed in the time-domain. Apodization is performed by speeding up, or slowing down the rate of decay of time-domain exponential functions. This is accomplished by multiplying the time-domain signal by another function. This technique allows the improvement of signal-to-noise ratio in exchange for the reduction of signal resolution (or visa versa).

Although many of these signal processing techniques have enjoyed success in specific applications, they remain solutions to specific types of aberrations. In order to produce an overall "clean" spectrum, one needs to use several of these methods to eliminate the aberrations present in a real spectrum. Furthermore, most of these techniques produce side effects that are magnified when improperly processed by a second signal processing algorithm. Furthermore, after the initial signal processing steps have been taken, the task of identifying the processed spectrum remains. This is not a trivial task as many times the quality of the processed spectrum remains poor, requiring a sophisticated identification system.

In this paper we show that instead of eliminating all the present aberrations by a signal processing procedure as a preprocessor, it is possible to eliminate some of them in the processing step, and some in the actual identification step. Here, we show that an adaptive identification system can learn to effectively ignore some forms of aberrations.

**Xyloglucan Oligosaccharides.** Complex carbohydrates are important biomolecules that play a role in many biological functions such as providing physical strength (connective tissue in animals and woody tissue in plants) and as a source of energy reserves (glycogen in animals and starch in plants). These molecules are also known to be directly and widely involved in biological recognition and regulatory processes in normal growth and development as well as in disease processes. The recent discovery of the role of complex carbohydrates in disease processes, and therefore drug development, among others has triggered a large number of studies in order to better understand the role of abnormal (structurally altered) complex carbohydrates in disease development. For this reason, an automated identification system of complex carbohydrates can eliminate the many man-hours wasted in duplicated efforts in structural characterization of known carbohydrates.

A specific group of these molecules from plant cell wall are called xyloglucan oligosaccharides. The $^1$H-NMR spectra of these molecules resemble each other to a great degree, and the experiments in developing an automated identification system for these spectra is a good indicator for the success of such future projects for automated identification of other molecules.

## 2. Method:

Two pattern recognition techniques were studied in this project, namely Bayesian classification [8], and artificial neural networks [9]. Multidimensional Parzen density estimation [10] was used to estimate the *a priori* probability density functions required for Bayes classification. For the ANN experiments, a feed-forward, 2-stage network trained with back-propagation learning algorithm was used to produce an identification system. Both identification systems were built using 30 spectra representing 30 unique xyloglucan oligosaccharides (training set), and tested with 30 newly recorded spectra of the same oligosaccharides in addition to 45 $^1$H-NMR spectra of complex carbohydrates other than xyloglucans. Each spectrum contained 5000 points representing the region between 1.0-5.5 ppm (parts per million). Five percent normally distributed noise was dynamically added to the spectra at the beginning of each ANN training epoch to prevent memorization. Same amount of noise was introduced to build a large database of spectra required by Parzen density estimation in order to accurately estimate the multidimensional densities. The optimal network configuration for the ANN was found to be 5000 input, 12 hidden, and 30 output neurons.

A preprocessing step is also implemented and kept constant for both identification techniques. This preprocessing step is comprised of several signal processing techniques that are intended to eliminate certain, but not all, aberrations present in $^1$H-NMR spectra of xyloglucan oligosaccharides. The preprocessing step includes, interpolation, a running window low pass filter for high frequency noise reduction, a ¾ scaling mechanism based on bin analysis for reducing the effects of sample concentration on signal strength, and a piecewise linear baseline correction routine.

## 3. Results:

The performance of the ANN was compared to that of a multidimensional Bayesian classifier. Table 1, shows the results of the first set of experiments. As can be seen, the performance of both methods was good during training. Although, Bayes' classifier misclassified one of the 30 xyloglucans from the training set. The two methods were tested with two testing sets. Testing set 1 included 30 new spectra of the same xyloglucans. Testing set 2, included 45 spectra of some carbohydrates other than xyloglucans. Testing set 2 was specifically designed to test the models for false positive errors. As it can be seen from Table 1, The ANN model performed better for all three data sets. However, both models need improvements to avoid false positives. The ANN model reported 4 false positives, while the Bayes' classifier reported 9. For testing set 2, the correct classification was considered to be a "no hit".

**Table 1. Number of correctly identified complex carbohydrates by the two methods.**

| Classification Method | Training Set | Testing Set 1 | Testing Set 2 |
|---|---|---|---|
| Artificial Neural Network | 30 | 30 | 41 |
| Parzen density estimation / Bayesian Classification | 29 | 28 | 36 |

A second set of experiments were conducted to test both models' noise tolerance. Three new testing sets were prepared from the original testing set 1. Each of the new sets contained the original testing spectra perturbed with 5%, 10%, and 15% white noise respectively. As it can be seen from Table 2, the performance of neither model degraded when 5% noise was added. The performance of the Bayes' classifier was even improved slightly. We hypothesize that this is due to the fact that both models were built with spectra that were 5% perturbed. We propose that the models have learned to filter out that level of noise. However, when the noise level was increased, the performance of both models degraded. This was especially evident for the Bayes' classifier. With 15% white noise, the performance of this model degraded to 18 correct identifications out of 30 carbohydrates.

**Table 2.** Number of correctly identified complex carbohydrates in presence of white noise.

| Classification Method | Testing Set 1 + 5% white noise | Testing Set 1 + 10% white noise | Testing Set 1 + 15% white noise |
|---|---|---|---|
| Artificial Neural Network | 30 | 28 | 23 |
| Parzen density estimation / Bayesian Classification | 29 | 27 | 18 |

## 4. New Aspects:

Separation of xyloglucan oligosaccharides based on their $^1$H-NMR spectra recorded on any 500 MHz NMR instrument is a non-linearly separable task. To the best knowledge of the authors, this is the first identification system for this group of molecules that is instrument insensitive. The system has been implemented on the web (http://www.ccrc.uga.edu/web/ccrcnet/ParsKimi.html), and already has been used by scientists around the world to identified over 250 spectra submitted via the world wide web.

## 5. Conclusions:

Xyloglucan oligosaccharides are a group of closely related plant cell wall complex carbohydrates whose spectra resemble each other to a great degree. The lack of a large number of clean $^1$H-NMR spectra of these structures has prevented the building of an accurate statistical model to identify these structures. For instance, Bayes' classification in combination with multidimensional Parzen density estimation did not perform well mainly due to a very sparse input space, and therefore, the failure to accurately estimate the distribution functions. We have developed an artificial neural network system that can successfully distinguish between the $^1$H-NMR spectra of these molecules. Furthermore, this model has not exhibited any instrument dependent sensitivity.

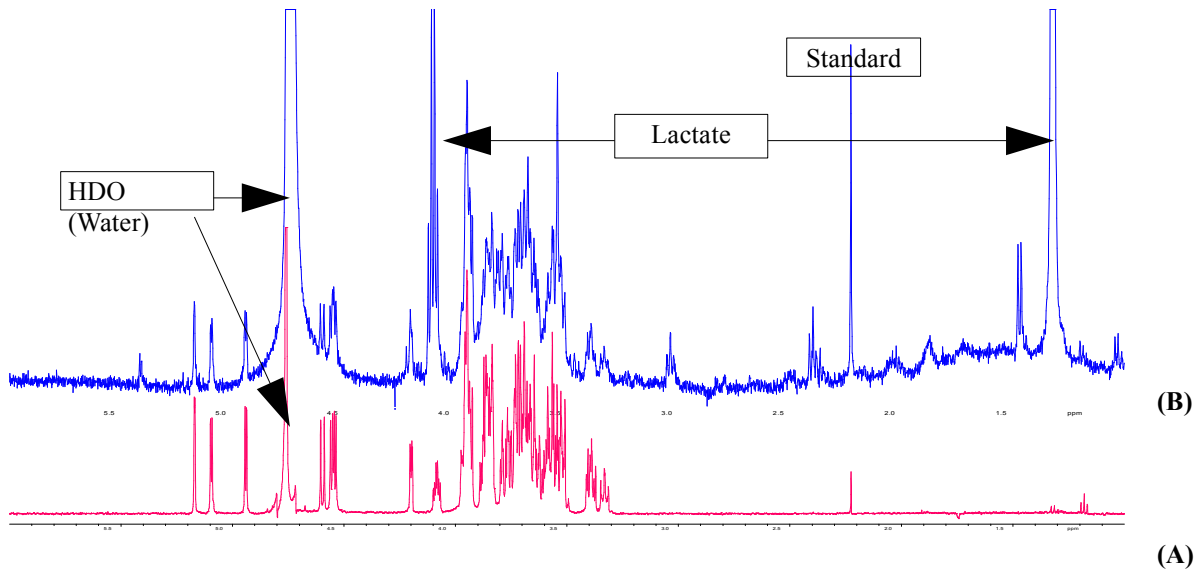

**Figure 1.** (A) A high quality $^1$H-NMR spectrum of a xyloglucan. (B) A poor quality $^1$H-NMR spectrum of the same oligosaccharide, with baseline drift, noise, negative signals, and large contaminant and standard signals.